\documentclass[12pt,a4paper]{article}
\makeatletter
\def\eqnarray{%
\stepcounter{equation}%
\let\@currentlabel=\theequation
\global\@eqnswtrue
\global\@eqcnt\z@
\tabskip\@centering
\let\\=\@eqncr
$$\halign to \displaywidth\bgroup\@eqnsel\hskip\@centering
$\displaystyle\tabskip\z@{##}$&\global\@eqcnt\@ne
\hfil$\displaystyle{{}##{}}$\hfil
&\global\@eqcnt\tw@$\displaystyle\tabskip\z@{##}$\hfil
\tabskip\@centering&\llap{##}\tabskip\z@\cr}
\makeatother
\newcommand{\kansu}[2]{{{#1}\!\left({#2}\right)}}
\newcommand{\ket}[1]{{\vert{#1}\rangle}}
\newcommand{\bra}[1]{{\langle{#1}\vert}}
\newcommand{\kett}[2]{{\vert{#1,#2}\rangle}}
\newcommand{\calh}{{\cal H}}
\newcommand{\calm}{{\cal M}}
\newcommand{\cala}{{\cal A}}
\newcommand{\calf}{{\cal F}}

\newcommand{\fukuso}{{\mathbf C}}
\newcommand{\futon}{{\bf N}}

\newcommand{\stm}{{St_m}}
\newcommand{\grm}{{Gr_m}}
\newcommand{\eem}{{E_m}}

\newcommand{\xizeta}{{\vert\xi\vert}}
\newcommand{\zezeta}{{\vert\zeta\vert}}
\newcommand{\lam}{{\bf \lambda}}
\newcommand{\slam}{{\bf \lambda_0}}

\begin{document}

\title{\sl Mathematical Foundations of Holonomic \\
           Quantum Computer}
\author{
  Kazuyuki FUJII
  \thanks{E-mail address : fujii@math.yokohama-cu.ac.jp }\\
  Department of Mathematical Sciences\\
  Yokohama City University\\
  Yokohama, 236-0027\\
  Japan
  }
\date{}
\maketitle\thispagestyle{empty}
%
%
%  gaiyou
%
%
\begin{abstract}
  We make a brief review of (optical) Holonomic Quantum Computer 
  (or Computation) proposed by Zanardi and Rasetti (quant--ph 9904011) 
  and Pachos and Chountasis (quant--ph 9912093), and give a 
  mathematical reinforcement to their works.
\end{abstract}

\newpage

%
%
%     Honbun
%
%

\section{Introduction}

Quantum Computer is a very attractive and challenging object for New 
Science.

After the breakthrough by P. Shor \cite{PS} there has been remarkable
progress in Quantum Computer or Computation (QC briefly).
This discovery had a great influence on scientists. This drived not only 
theoreticians to finding other quantum algorithms, but also 
experimentalists to building quantum computers.
See \cite{AS} and \cite{RP} in outline. \cite{LPS} is also very useful.

On the other hand, Gauge Theories are widely recognized as the basis in 
quantum field theories.
Therefore it is very natural to intend to include gauge theories 
in QC $\cdots$ a construction of ``gauge theoretical'' quantum computation
or of ``geometric'' quantum computation in our terminology. 
The merit of geometric method of QC is strong against the influence 
from the environment. See \cite{JVEC}. 

Zanardi and Rasetti in \cite{ZR} and \cite{PZR} proposed such an idea 
using non-abelian Berry phase (quantum holonomy). See also  
\cite{JP} and \cite{AYK} as another geometric models.
In their model a Hamiltonian (including some parameters) must be
degenerated because an adiabatic connection is introduced using
this degeneracy \cite{SW}. 
In other words, a quantum computational bundle on some parameter space 
(see \cite{ZR}) is introduced due to this degeneracy.

They gave a few simple but interesting examples to explain their idea. We 
believe that these examples will become important in the near future. 
But their works (\cite{ZR}, \cite{PZR} and \cite{PC}) are a bit coarse in 
the mathematical point of view. 
Therefore in this paper we give a mathematical reinforcement to 
them. See \cite{KF2} and also \cite{KF3} as a further generalization.

It is not easy to predict the future of geometric quantum computation.
However it is an arena worth challenging for mathematical physicists.

%\vspace{5mm}
\section{Mathematical Foundation of Quantum Holonomy}

We start with mathematical preliminaries.
Let $\calh$ be a separable Hilbert space over $\fukuso$.
For $m\in{\bf N}$, we set
\begin{equation}
  \label{eq:stmh}
  \kansu{\stm}{\cal H}
  \equiv
  \left\{
    V=\left( v_1,\cdots,v_m\right)
    \in
    \calh\times\cdots\times\calh\vert V^\dagger V=1_m\right\}\ ,
\end{equation}
where $1_m$ is a unit matrix in $\kansu{M}{m,\fukuso}$.
This is called a (universal) Stiefel manifold.
Note that the unitary group $U(m)$ acts on $\kansu{\stm}{\calh}$
from the right:
\begin{equation}
  \label{eq:stmsha}
  \kansu{\stm}{\calh}\times\kansu{U}{m}
  \rightarrow
  \kansu{\stm}{\calh}: \left( V,a\right)\mapsto Va.
\end{equation}
Next we define a (universal) Grassmann manifold
\begin{equation}
  \kansu{\grm}{\calh}
  \equiv
  \left\{
    X\in\kansu{M}{\calh}\vert
    X^2=X, X^\dagger=X\  \mathrm{and}\  \mathrm{tr}X=m\right\}\ ,
\end{equation}
where $M(\calh)$ denotes a space of all bounded linear operators on $\calh$.
Then we have a projection
\begin{equation}
  \label{eq:piteigi}
  \pi : \kansu{\stm}{\calh}\rightarrow\kansu{\grm}{\calh}\ ,
  \quad \kansu{\pi}{V}\equiv VV^\dagger\ ,
\end{equation}
compatible with the action (\ref{eq:stmsha}) 
($\kansu{\pi}{Va}=Va(Va)^\dagger=Vaa^\dagger V^\dagger=VV^\dagger=
\kansu{\pi}{V}$).

Now the set
\begin{equation}
  \label{eq:principal}
  \left\{
    \kansu{U}{m}, \kansu{\stm}{\calh}, \pi, \kansu{\grm}{\calh}
  \right\}\ ,
\end{equation}
is called a (universal) principal $U(m)$ bundle, 
see \cite{MN} and \cite{KF1}.\quad We set
\begin{equation}
  \label{eq:emh}
  \kansu{\eem}{\cal H}
  \equiv
  \left\{
    \left(X,v\right)
    \in
    \kansu{\grm}{\calh}\times\calh \vert Xv=v \right\}\ .
\end{equation}
Then we have also a projection 
\begin{equation}
  \label{eq:piemgrm}
  \pi : \kansu{\eem}{\calh}\rightarrow\kansu{\grm}{\calh}\ ,
  \quad \kansu{\pi}{\left(X,v\right)}\equiv X\ .
\end{equation}
The set
\begin{equation}
  \label{eq:universal}
  \left\{
    \fukuso^m, \kansu{\eem}{\calh}, \pi, \kansu{\grm}{\calh}
  \right\}\ ,
\end{equation}
is called a (universal) $m$-th vector bundle. This vector bundle is 
one associated with the principal $U(m)$ bundle (\ref{eq:principal})
.

Next let $M$ be a finite or infinite dimensional differentiable manifold 
and the map $P : M \rightarrow \kansu{\grm}{\calh}$ be given (called a 
projector). Using this $P$ we can make 
the bundles (\ref{eq:principal}) and (\ref{eq:universal}) pullback 
over $M$ :
\begin{eqnarray}
  \label{eq:hikimodoshi1}
  &&\left\{\kansu{U}{m},\widetilde{St}, \pi_{\widetilde{St}}, M\right\}
  \equiv
  P^*\left\{\kansu{U}{m}, \kansu{\stm}{\calh}, \pi, 
  \kansu{\grm}{\calh}\right\}
  \ , \\
  \label{eq:hikimodoshi2}
  &&\left\{\fukuso^m,\widetilde{E}, \pi_{\widetilde{E}}, M\right\}
  \equiv
  P^*\left\{\fukuso^m, \kansu{\eem}{\calh}, \pi, \kansu{\grm}{\calh}\right\}
  \ ,
\end{eqnarray}
see \cite{MN}. (\ref{eq:hikimodoshi2}) is of course a vector bundle 
associated with (\ref{eq:hikimodoshi1}).

Let $\calm$ be a parameter space and we denote by $\lam$ its element. 
Let $\slam$ be a fixed reference point of $\calm$. Let $H_\lam$ be 
a family of Hamiltonians parametrized by $\calm$ which act on a Fock space 
$\calh$. We set $H_0$ = $H_\slam$ for simplicity and assume that this has 
a $m$-fold degenerate vacuum :
\begin{equation}
  H_{0}v_j = \mathbf{0},\quad j = 1 \sim m. 
\end{equation}
These $v_j$'s form a $m$-dimensional vector space. We may assume that 
$\langle v_{i}\vert v_{j}\rangle = \delta_{ij}$. Then $\left(v_1,\cdots,v_m
\right) \in \kansu{\stm}{\calh}$ and 
\[
  F_0 \equiv \left\{\sum_{j=1}^{m}x_{j}v_{j}\vert x_j \in \fukuso \right\} 
  \cong \fukuso^m.
\]
Namely, $F_0$ is a vector space associated with o.n.basis 
$\left(v_1,\cdots,v_m\right)$.

Next we assume for simplicity 
that a family of unitary operators parametrized by $\calm$
\begin{equation}
  \label{eq:ufamily} 
  W : \calm \rightarrow U(\calh),\quad W(\slam) = {\rm id}.
\end{equation}
is given and $H_{\lam}$ above is given by the following isospectral family
\begin{equation}
 H_{\lam} \equiv W(\lam)H_0 W(\lam)^{-1}.
\end{equation}
In this case there is no level crossing of eigenvalues. Making use of 
$W(\lam)$ we can define a projector
\begin{equation}
  \label{eq:pfamily}
 P : \calm \rightarrow \kansu{\grm}{\calh}, \quad 
 P(\lam) \equiv
  W(\lam) \left(\sum^{m}_{j=1}v_{j}v_{j}^{\dagger}\right)W(\lam)^{-1}
\end{equation}
and have the pullback bundles over $\calm$
\begin{equation}
  \label{eq:target}
 \left\{\kansu{U}{m},\widetilde{St}, \pi_{\widetilde{St}}, \calm\right\},\quad 
 \left\{\fukuso^m,\widetilde{E}, \pi_{\widetilde{E}}, \calm\right\}.
\end{equation}

For the later we set
\begin{equation}
  \label{eq:vacuum}
 \ket{vac} = \left(v_1,\cdots,v_m\right).
\end{equation}
In this case a canonical connection form $\cala$ of 
$\left\{\kansu{U}{m},\widetilde{St}, \pi_{\widetilde{St}}, \calm\right\}$ is 
given by 
\begin{equation}
  \label{eq:cform}
 \cala = \bra{vac}W(\lam)^{-1}d W(\lam)\ket{vac},
\end{equation}
where $d$ is a differential form on $\calm$, and its curvature form by
\begin{equation}
  \label{eq:curvature}
  \calf \equiv d\cala+\cala\wedge\cala,
\end{equation}
see \cite{SW} and \cite{MN}.

Let $\gamma$ be a loop in $\calm$ at $\slam$., $\gamma : [0,1] 
\rightarrow \calm, \gamma(0) = \gamma(1)$. For this $\gamma$ a holonomy 
operator $\Gamma_{\cala}$ is defined :
\begin{equation}
  \label{eq:holonomy}
  \Gamma_{\cala}(\gamma) = {\cal P}exp\left\{\oint_{\gamma}\cala\right\} 
  \in \kansu{U}{m},
\end{equation}
where ${\cal P}$ means path-ordered. This acts on the fiber $F_0$ at 
$\slam$ of the vector bundle 
$\left\{\fukuso^m,\widetilde{E}, \pi_{\widetilde{E}}, M\right\}$ as follows :
${\textbf x} \rightarrow \Gamma_{\cala}(\gamma){\textbf x}$.\quad 
The holonomy group $Hol(\cala)$ is in general subgroup of $\kansu{U}{m}$ 
. In the case of $Hol(\cala) = \kansu{U}{m}$,   $\cala$ is called 
irreducible, see \cite{MN}.

In the Holonomic Quantum Computer we take  
\begin{eqnarray}
  \label{eq:information}
  &&{\rm Encoding\ of\ Information} \Longrightarrow {\textbf x} \in F_0 , 
  \nonumber \\
  &&{\rm Processing\ of\ Information} \Longrightarrow \Gamma_{\cala}(\gamma) : 
  {\textbf x} \rightarrow \Gamma_{\cala}(\gamma){\textbf x}.
\end{eqnarray}

%\vspace{5mm}
\section{Holonomic Quantum Computation}

We apply the results of last section to Quantum Optics and discuss about 
(optical) Holonomic Quantum Computation proposed by \cite{ZR} and 
\cite{PC}.

Let $a(a^\dagger)$ be the annihilation (creation) operator of the harmonic 
oscillator.
If we set $N\equiv a^\dagger a$ (:\ number operator), then
\begin{equation}
  [N,a^\dagger]=a^\dagger\ ,\
  [N,a]=-a\ ,\
  [a,a^\dagger]=1\ .
\end{equation}
Let $\calh$ be a Fock space generated by $a$ and $a^\dagger$, and
$\{\ket{n}\vert n\in\futon\cup\{0\}\}$ be its basis.
The actions of $a$ and $a^\dagger$ on $\calh$ are given by
\begin{equation}
  \label{eq:shoukou}
  a\ket{n} = \sqrt{n}\ket{n-1}\ ,\
  a^\dagger\ket{n} = \sqrt{n+1}\ket{n+1}\ ,
\end{equation}
where $\ket{0}$ is a vacuum ($a\ket{0}=0$).

Next we consider the system of two-harmonic oscillators. If we set
\begin{equation}
  \label{eq:twosystem}
  a_1 = a \otimes 1,\  a_1^{\dagger} = a^{\dagger} \otimes 1;\ 
  a_2 = 1 \otimes a,\  a_2^{\dagger} = 1 \otimes a^{\dagger},
\end{equation}
then it is easy to see 
\begin{equation}
  \label{eq:relations}
 [a_i, a_j] = [a_i^{\dagger}, a_j^{\dagger}] = 0,\ 
 [a_i, a_j^{\dagger}] = \delta_{ij}, \quad i, j = 1, 2. 
\end{equation}
We also denote by $N_{i} = a_i^{\dagger}a_i$ number operators.

Now since we want to consider coherent states based on Lie algebras $su(2)$ 
and $su(1,1)$, we make use of Schwinger's boson method, see \cite{FKSF1}, 
\cite{FKSF2}. Namely if we set 
\begin{eqnarray}
  \label{eq:Jdaisu}
  {\rm [C]}&&\ su(2) :\quad
     J_+ = a_1^{\dagger}a_2,\ J_- = a_2^{\dagger}a_1,\ 
     J_3 = {1\over2}\left(a_1^{\dagger}a_1 - a_2^{\dagger}a_2\right), \\
  \label{eq:Kdaisu}
  {\rm [NC]}&&\ su(1,1) :\quad
     K_+ = a_1^{\dagger}a_2^{\dagger},\ K_- = a_2 a_1,\ 
     K_3 = {1\over2}\left(a_1^{\dagger}a_1 + a_2^{\dagger}a_2  + 1\right),
\end{eqnarray}
then we have
\begin{eqnarray}
  \label{eq:j-relation}
  {\rm [C]}&&\ su(2) :\quad
     [J_3, J_+] = J_+,\ [J_3, J_-] = - J_-,\ [J_+, J_-] = 2J_3, \\
  \label{eq:k-relation}
  {\rm [NC]}&&\ su(1,1) :\quad
     [K_3, K_+] = K_+,\ [K_3, K_-] = - K_-,\ [K_+, K_-] = -2K_3.
\end{eqnarray}

In the following we treat unitary coherent operators based on Lie algebras 
$su(2)$ and $su(1,1)$. 

\noindent{\bfseries Definition}\quad We set 
\begin{eqnarray}
  \label{eq:j-operator}
  {\rm [C]}&&\ 
U(\xi) = e^{\xi a_1^{\dagger}a_2 - \bar{\xi}a_2^{\dagger}a_1}\quad 
  {\rm for}\  \xi \in \fukuso , \\
  \label{eq:k-operator}
  {\rm [NC]}&&\ 
V(\zeta) = e^{\zeta a_1^{\dagger}a_2^{\dagger} - \bar{\zeta}a_2 a_1}\quad 
  {\rm for}\  \zeta \in \fukuso.
\end{eqnarray}
For the details of $U(\xi)$ and $ V(\zeta)$ see \cite{AP} and 
\cite{FKSF1}. For the latter convenience let us list well-known 
disentangling formulas.

\noindent{\bfseries Lemma 1}\quad We have
\begin{eqnarray}
  \label{eq:j-formula}
{\rm [C]}\ U(\xi) &=& e^{\eta a_1^{\dagger}a_2}
 e^{{\rm log}\left(1 + {\vert \eta \vert}^{2}\right)
    {1\over2}\left(a_1^{\dagger}a_1 - a_2^{\dagger}a_2\right)}
 e^{- \bar{\eta}a_2^{\dagger}a_1}, \quad 
   where\quad  \eta = \frac{\xi {\rm tan}{\vert \xi \vert}}
                           {{\vert \xi \vert}},  \\
  \label{eq:k-formula}
{\rm [NC]}\ V(\zeta) &=& e^{\kappa a_1^{\dagger}a_2^{\dagger}}
 e^{{\rm log}\left(1 - {\vert \kappa \vert}^{2}\right)
    {1\over2}\left(a_1^{\dagger}a_1 + a_2^{\dagger}a_2 + 1\right)}
 e^{- \bar{\kappa}a_2 a_1},\quad 
   where\quad  \kappa = \frac{\zeta {\rm tanh}{\vert \zeta \vert}}
                             {{\vert \zeta \vert}}. 
 \end{eqnarray}
As for a genelalization of these formulas see \cite{FS}.

Let $H_0$ be a Hamiltonian with nonlinear interaction produced by 
a Kerr medium., that is  $H_0 = \hbar {\rm X} N(N-1)$, where X is a 
certain constant, see \cite{PC}. The eigenvectors of $H_0$ corresponding 
to $0$ is $\left\{\ket{0},\ket{1}\right\}$, so its eigenspace is 
${\rm Vect}\left\{\ket{0},\ket{1}\right\} \cong \fukuso^2$. We correspond to 
$0 \rightarrow \ket{0},\ 1 \rightarrow \ket{1}$ for a generator of 
Boolean algebra $\left\{0, 1\right\}$. The space 
${\rm Vect}\left\{\ket{0},\ket{1}\right\}$ is called 1-qubit (quantum bit) 
space, see \cite{AS} or  \cite{RP}. 
Since we are considering the system of two particles, the Hamiltonian that 
we treat in the following is 
\begin{equation}
  \label{eq:hamiltonian}
  H_0 = \hbar {\rm X} N_{1}(N_{1}-1) + \hbar {\rm X} N_{2}(N_{2}-1).
\end{equation}
The eigenspace of $0$ of this Hamiltonian becomes therefore 
\begin{equation}
  \label{eq:eigenspace}
   F_0 = {\rm Vect}\left\{\ket{0},\ket{1}\right\}\otimes 
         {\rm Vect}\left\{\ket{0},\ket{1}\right\} 
         \cong \fukuso^2\otimes \fukuso^2.
\end{equation}
We denote the basis of $F_0$ as $\left\{\kett{0}{0},\kett{0}{1}, \kett{1}{0},
 \kett{1}{1}\right\}$ and set $\ket{vac} = \left(\kett{0}{0}, \kett{0}{1}, 
\kett{1}{0}, \kett{1}{1} \right)$.

Next we consider the following isospectral family of $H_0$ above :
\begin{eqnarray}
  \label{eq:twofamily}
   H_{(\xi,\zeta)}&=& W(\xi,\zeta)H_0 W(\xi,\zeta)^{-1},\\
  \label{eq:double}
   W(\xi,\zeta) &=&  U(\xi)V(\zeta) \in U(\calh\otimes \calh),\quad 
   W(0,0) = {\rm id}.
\end{eqnarray}
For this system let us calculate a connection form (\ref{eq:cform})
in the last section. For that we set 
\begin{equation}
  \label{eq:coefficients}
  A_{\xi} = \bra{vac}W(\xi,\zeta)^{-1}\frac{\partial}{\partial \xi}
            W(\xi,\zeta)\ket{vac}, \quad 
  A_{\zeta} = \bra{vac}W(\xi,\zeta)^{-1}\frac{\partial}{\partial \zeta}
            W(\xi,\zeta)\ket{vac}.
\end{equation}
Here remaking
\begin{eqnarray}
 W(\xi,\zeta)^{-1}\frac{\partial}{\partial \xi}W(\xi,\zeta) &=& 
 V(\xi)^{-1}\left\{U(\xi)^{-1}\frac{\partial}{\partial \xi}U(\xi)\right\}
 V(\xi), \nonumber \\
 W(\xi,\zeta)^{-1}\frac{\partial}{\partial \zeta}W(\xi,\zeta) &=& 
 V(\xi)^{-1}\frac{\partial}{\partial \zeta}V(\xi) \nonumber
\end{eqnarray}
and using Lemma 1,
 
\noindent{\bfseries Lemma 2}\quad we have
\begin{eqnarray}
 &&W^{-1}\frac{\partial}{\partial \xi}W   \nonumber \\
 &=& 
 {1\over2}\left(1+{\sin(2\xizeta)\over2\xizeta}\right)
 \left\{
 \cosh(2\zezeta)a_1^{\dagger}a_2 + 
 {\zeta\sinh(2\zezeta)\over2\zezeta}\left(a_1^{\dagger}\right)^2 +
 {{\bar \zeta}\sinh(2\zezeta)\over2\zezeta}\left(a_2\right)^2 
 \right\} \nonumber \\
 &+& {{\bar \xi}\over2\xizeta^{2}}\left(1-\cos(2\xizeta)\right)
  {1\over2}\left(a_1^{\dagger}a_1 -  a_2^{\dagger}a_2\right) \nonumber \\
 &+&{{\bar \xi}^{2}\over2\xizeta^{2}}
 \left(-1+{\sin(2\xizeta)\over2\xizeta}\right)
 \left\{
 \cosh(2\zezeta)a_2^{\dagger}a_1 + 
 {{\bar \zeta}\sinh(2\zezeta)\over2\zezeta}\left(a_1\right)^2 +
 {\zeta\sinh(2\zezeta)\over2\zezeta}\left(a_2^{\dagger}\right)^2 
 \right\}, \\
 &&W^{-1}\frac{\partial}{\partial \zeta}W   \nonumber \\
 &=& 
 {1\over2}\left(1+{\sinh(2\zezeta)\over2\zezeta}\right)
  a_1^{\dagger}a_2^{\dagger} 
  + {{\bar \zeta}\over2\zezeta^{2}}\left(-1+ \cosh(2\zezeta)\right)
  {1\over2}\left(a_1^{\dagger}a_1 + a_2^{\dagger}a_2 + 1\right) \nonumber \\
 &+&{{\bar \zeta}^{2}\over2\zezeta^{2}}
 \left(-1+{\sinh(2\zezeta)\over2\xizeta}\right)a_1a_2.
\end{eqnarray}
From this lemma it is easy to calculate $A_{\xi}$ and $A_{\zeta}$.
Before stating the result let us prepare some notations.
\begin{equation}
{\widehat E} = 
\left(
  \begin{array}{cccc}
    0& 0& 0& 0 \\
    0& 0& 1& 0 \\
    0& 0& 0& 0 \\
    0& 0& 0& 0 
  \end{array}
\right), 
{\widehat F} = 
\left(
  \begin{array}{cccc}
    0& 0& 0& 0 \\
    0& 0& 0& 0 \\
    0& 1& 0& 0 \\
    0& 0& 0& 0 
  \end{array}
\right), 
{\widehat H} = 
\left(
  \begin{array}{cccc}
    0& 0& 0& 0 \\
    0& {1\over2}& 0& 0 \\
    0& 0& -{1\over2}& 0 \\
    0& 0& 0& 0 
  \end{array}
\right).
\end{equation}
\begin{equation}
{\widehat A} = 
\left(
  \begin{array}{cccc}
    0& 0& 0& 1 \\
    0& 0& 0& 0 \\
    0& 0& 0& 0 \\
    0& 0& 0& 0 
  \end{array}
\right), 
{\widehat C} = 
\left(
  \begin{array}{cccc}
    0& 0& 0& 0 \\
    0& 0& 0& 0 \\
    0& 0& 0& 0 \\
    1& 0& 0& 0 
  \end{array}
\right), 
{\widehat B} = 
\left(
  \begin{array}{cccc}
    {1\over2}& 0& 0& 0 \\
    0& 1& 0& 0 \\
    0& 0& 1& 0 \\
    0& 0& 0& {3\over2}
  \end{array}
\right).
\end{equation}

\noindent{\bfseries Proposition 3}\quad We have
\begin{eqnarray}
 A_{\xi}&=& 
 {1\over2}\left(1+{\sin(2\xizeta)\over2\xizeta}\right)\cosh(2\zezeta)
 {\widehat F} 
  - {{\bar \xi}\over2\xizeta^{2}}\left(1-\cos(2\xizeta)\right){\widehat H} 
 \nonumber \\
  &&+ {{\bar \xi}^{2}\over2\xizeta^{2}}
 \left(-1+{\sin(2\xizeta)\over2\xizeta}\right)\cosh(2\zezeta){\widehat E}, \\
A_{\zeta} &=&  
 {1\over2}\left(1+{\sinh(2\zezeta)\over2\zezeta}\right){\widehat C}
  + 
 {{\bar \zeta}\over2\zezeta^{2}}\left(-1 + \cosh(2\zezeta)\right){\widehat B}
 \nonumber \\ 
  &&+ {{\bar \zeta}^{2}\over2\zezeta^{2}}
 \left(-1+{\sinh(2\zezeta)\over2\zezeta}\right){\widehat A}.
 \end{eqnarray}
Since the connection form $\cala$ is anti-hermitian 
($\cala^{\dagger}=-\cala$), 
it can be written as
\begin{equation}
%  \label{eq:calaa}
  \cala =
  A_\xi d\xi + A_\zeta d\zeta - A_\xi^{\dagger} d{\bar \xi}
  -A_\zeta^{\dagger} d{\bar \zeta}\ ,
\end{equation}
so that it's curvature form $\calf = d\cala + \cala\wedge\cala$ becomes
\begin{eqnarray}
  \label{eq:explicit}
  \calf
  &=&
  \left(
    \partial_\xi A_\zeta-\partial_\zeta A_\xi
    +[A_\xi,A_\zeta]
  \right) d\xi\wedge d\zeta
  \nonumber\\
  &&
  -\left(
    \partial_\xi A_\xi^\dagger+\partial_{\bar\xi}A_\xi
    +[A_\xi,A_\xi^\dagger]
  \right) d\xi\wedge d\bar\xi
  \nonumber\\
  &&
   -\left(
    \partial_\xi A_\zeta^\dagger+\partial_{\bar\zeta}A_\xi
    +[A_\xi,A_\zeta^\dagger]
  \right) d\xi\wedge d\bar\zeta
  \nonumber\\
  &&-\left(
    \partial_\zeta A_\xi^\dagger+\partial_{\bar\xi}A_\zeta
    +[A_\zeta,A_\xi^\dagger]
  \right) d\zeta\wedge d\bar\xi
  \nonumber\\
  &&-\left(
    \partial_\zeta A_\zeta^\dagger+\partial_{\bar\zeta}A_\zeta
    +[A_\zeta,A_\zeta^\dagger]
  \right) d\zeta\wedge d\bar\zeta
  \nonumber\\
  &&-\left(
    \partial_{\bar\xi}A_\zeta^\dagger-\partial_{\bar\zeta}A_\xi^\dagger
    +[A_\zeta^\dagger,A_\xi^\dagger]
  \right) d\bar\xi\wedge d\bar\zeta .
\end{eqnarray}

Now we state our main result.

\noindent{\bfseries Theorem 4}
\begin{eqnarray}
  \label{eq:main-result}
&& \calf = \nonumber\\
  &&
  -\Bigg\{
 \left(1+{\sin(2\xizeta)\over2\xizeta}\right)
 {{\bar \zeta}\sinh(2\zezeta)\over2\zezeta}  
 {\widehat F} + 
 {{\bar \xi}^{2}\over\xizeta^{2}}
 \left(-1+{\sin(2\xizeta)\over2\xizeta}\right)
 {{\bar \zeta}\sinh(2\zezeta)\over2\zezeta}  
 {\widehat E}
     \Bigg\}
  d\xi\wedge d\zeta
  \nonumber\\
  &&
  -\Bigg\{
  {\xi\over\xizeta^{2}}
  \left(-1+\cos(2\xizeta)\right)\cosh(2\zezeta){\widehat F}
  -{\sin(2\xizeta)\over\xizeta}
   \left(1+\cosh^2(2\zezeta)\right){\widehat H} \nonumber\\
 &&\ \ \ \ \ +{{\bar\xi}\over\xizeta^{2}}
 \left(-1+\cos(2\xizeta)\right)\cosh(2\zezeta){\widehat E}
     \Bigg\}
  d\xi\wedge d\bar\xi
  \nonumber\\
  &&
  -\Bigg\{
 \left(1+{\sin(2\xizeta)\over2\xizeta}\right)
 {\zeta\sinh(2\zezeta)\over2\zezeta}{\widehat F}  
 + 
 {{\bar\xi}^{2}\over\xizeta^{2}}
 \left(-1+{\sin(2\xizeta)\over2\xizeta}\right)
 {\zeta\sinh(2\zezeta)\over2\zezeta}{\widehat E}
     \Bigg\}
  d\xi\wedge d\bar\zeta
 \nonumber\\
  &&
  -\Bigg\{
 \left(1+{\sin(2\xizeta)\over2\xizeta}\right)
 {{\bar\zeta}\sinh(2\zezeta)\over2\zezeta}{\widehat E}  
 +
 {\xi^{2}\over\xizeta^{2}}
 \left(-1+{\sin(2\xizeta)\over2\xizeta}\right)
 {{\bar\zeta}\sinh(2\zezeta)\over2\zezeta}{\widehat F}
     \Bigg\}
  d\zeta\wedge d\bar\xi
 \nonumber\\
  &&
 - {\sinh(2\zezeta)\over\zezeta}\left(2{\widehat B}-{\textbf 1}_4\right)
  d\zeta\wedge d\bar\zeta
  \nonumber\\
  &&
  +\Bigg\{
 \left(1+{\sin(2\xizeta)\over2\xizeta}\right)
 {\zeta\sinh(2\zezeta)\over2\zezeta}{\widehat E}  
 +
 {\xi^{2}\over\xizeta^{2}}
 \left(-1+{\sin(2\xizeta)\over2\xizeta}\right)
 {\zeta\sinh(2\zezeta)\over2\zezeta}{\widehat F}
     \Bigg\}
  d\bar\xi\wedge d\bar\zeta.
\end{eqnarray}
From this and the theorem of Ambrose--Singer (see \cite{MN}) 
it is easy to see that 

\noindent{\bfseries Corollary 5}
\begin{equation}
  Hol(\cala) = SU(2)\times U(1)\ \subset \ U(4).
\end{equation}
Therefore $\cala$ is not irreducible.

%\noindent{\em Acknowledgment.}\\
%The author wishes to thank Dr. K. Funahashi for his helpful comments and  
%suggestions. 

%%%%%%%%%%%%
%References%
%%%%%%%%%%%%

\end{document}